\documentclass[conference]{IEEEtran}
\IEEEoverridecommandlockouts
\usepackage{cite}
\usepackage{amsmath,amssymb,amsfonts}
\usepackage{algorithmic}
\usepackage{graphicx}
\graphicspath{ {Figures/} }
\usepackage{textcomp}
\usepackage{xcolor}
\usepackage{hyperref}
\def\BibTeX{{\rm B\kern-.05em{\sc i\kern-.025em b}\kern-.08em
    T\kern-.1667em\lower.7ex\hbox{E}\kern-.125emX}}
\begin{document}

\title{Poster Abstract: Towards Scalable and Trustworthy Decentralized Collaborative Intrusion Detection System for IoT}

\author{
\IEEEauthorblockN{
    Guntur Dharma Putra\IEEEauthorrefmark{1},
    Volkan Dedeoglu\IEEEauthorrefmark{2},
    Salil S. Kanhere\IEEEauthorrefmark{1},
    and Raja Jurdak\IEEEauthorrefmark{3}
}
\IEEEauthorblockA{
    \IEEEauthorrefmark{1}UNSW, Sydney
    \IEEEauthorrefmark{2}CSIRO Data61, Brisbane
    \IEEEauthorrefmark{3}QUT, Brisbane \\
    \{gdputra, salil.kanhere\}@unsw.edu.au, volkan.dedeoglu@data61.csiro.au, r.jurdak@qut.edu.au
}
}

\maketitle

\begin{abstract}
An Intrusion Detection System (IDS) aims to alert users of incoming attacks by deploying a detector that monitors network traffic continuously. As an effort to increase detection capabilities, a set of independent IDS detectors typically work collaboratively to build intelligence of holistic network representation, which is referred to as Collaborative Intrusion Detection System (CIDS). However, developing an effective CIDS, particularly for the IoT ecosystem raises several challenges. Recent trends and advances in blockchain technology, which provides assurance in distributed trust and secure immutable storage, may contribute towards the design of effective CIDS. In this poster abstract, we present our ongoing work on a decentralized CIDS for IoT, which is based on blockchain technology.
We propose an architecture that provides accountable trust establishment, which promotes incentives and penalties, and scalable intrusion information storage by exchanging bloom filters.
We are currently implementing a proof-of-concept of our modular architecture in a local test-bed and evaluate its effectiveness in detecting common attacks in IoT networks and the associated overhead.
\end{abstract}


\section{Introduction}
An Intrusion Detection System (IDS) is deployed as an indispensable defense mechanism, which comprised of monitors across the network, to identify and alert users about incoming attacks. Generally, a monitor identifies incoming threats either by matching observed events with known signature of intrusions (signature-based) or by finding deviations in current network traffic conditions and regular conditions by means of machine learning techniques (anomaly-based). Sensed threats are reported to the user as security alarms.

The prevalence of IoT implementation across diverse areas has expanded the attack surface and raised more security concerns, especially in protecting devices from various threats~\cite{Anthi2019}.
To improve the accuracy of threat detection, researchers proposed combination of individual IDSs to work in a collaborative fashion, referred as Collaborative-IDS (CIDS). In this scheme, a set of independent IDS monitors are encouraged to exchange intrusion information, e.g., alarms and attack signatures, resulting in a holistic representation of the recent network conditions.

While CIDS enhances the security, several inherent challenges of designing CIDS for IoT remain unsolved. First, IDS nodes need to trust one another to collaboratively exchange intrusion related information, especially if monitors are under the control of different owners.
There is a need for fair and transparent trust mechanisms that encourage incentives for trustworthy contributions and impose penalties for fraudulent manipulations.
Second, there is a need to design scalable mechanisms for storing a large amount of detection information that the individual IDS's within a CIDS may generate.
Third, the design of CIDS should also consider unique characteristics of the IoT network, such as the
highly dynamic network architecture and the constrained nature of IoT nodes.

\begin{figure}[!t]
\centering
\includegraphics[width=0.45\textwidth]{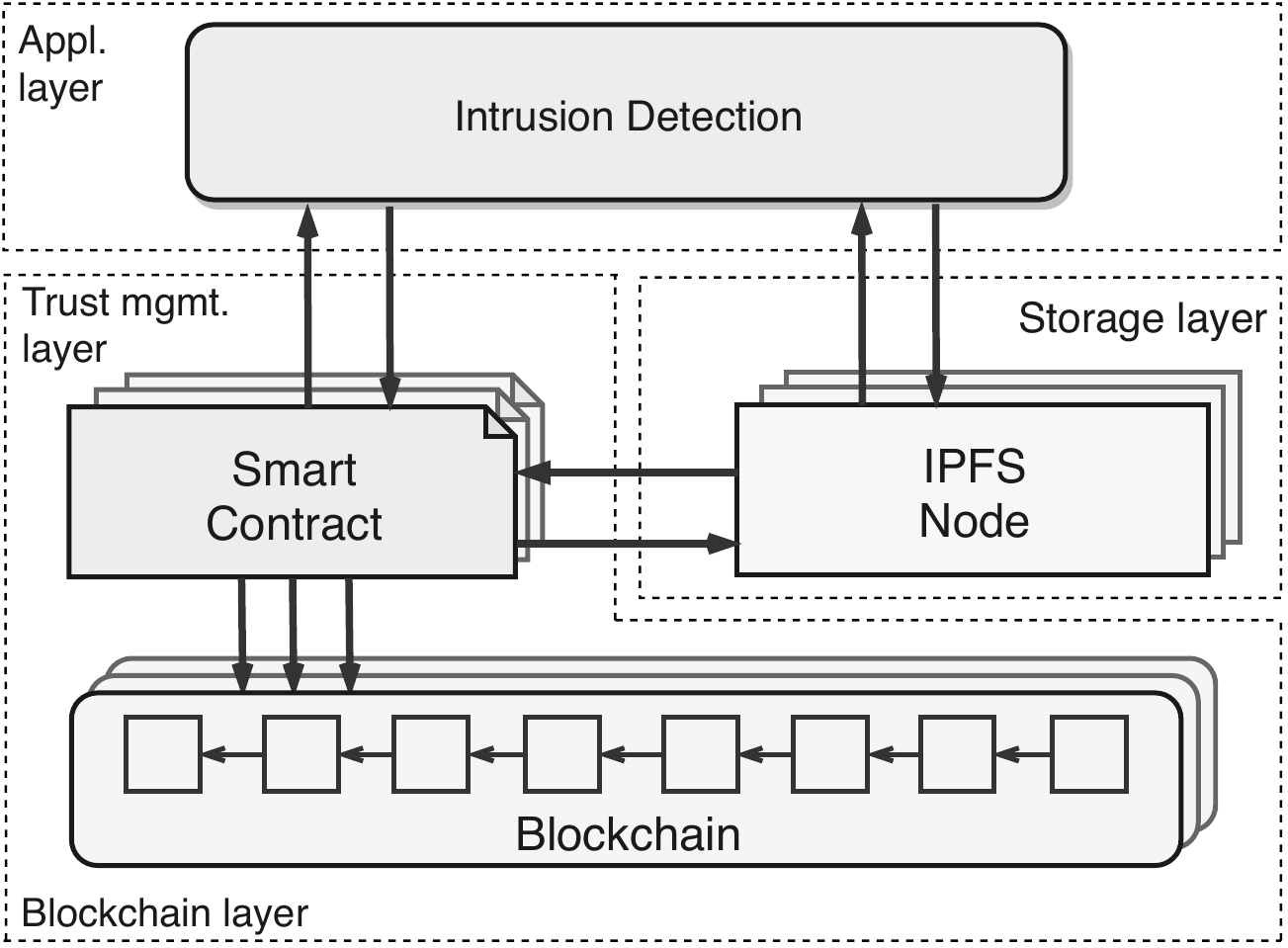}
\caption{Architecture of proposed decentralized CIDS.}
\label{fig:architecture-diagram}
\end{figure}

Recent research in blockchain-based technologies has shown its applicability in diverse domains, which solves challenges of decentralization of trust and reliable data storage in parallel and distributed systems. The decentralized nature of blockchain, that replaces centralized processing with secure collaborative processing, is a good fit for solving the inherent challenges in CIDS. Several works have proposed blockchain-based CIDS, including anomaly-based~\cite{Golomb2018} and signature-based~\cite{Tug2018} approaches, which essentially aim to progressively build trusted anomaly models and signature databases stored securely in blockchains. However, these works overlooked the aforementioned challenges, such as accountable trust establishment and scalability.

In this poster abstract, we present our ongoing work on decentralized CIDS for IoT.
Our blockchain-based design i) establishes distributed trust between participants, with transparent and accountable mechanisms that incentivize decent contributions, ii) allows compact representation of detection related information by exchanging bloom filters, which improves scalability, and iii) enables secure and efficient storage of the detection models and signature database by dividing meta-data and detection models to different storage media. We outline our architectural design in Fig.~\ref{fig:architecture-diagram}.

\section{Ongoing work}
We describe our ongoing work with regard to the architecture design and the evaluation strategy.

    \subsection{Architecture Design}
    As depicted in Fig.~\ref{fig:architecture-diagram}, our architecture consists of four interconnected layers, namely application, storage, trust management, and blockchain layers. The application layer is comprised of an IDS module which trains and runs detection models and collects intrusion signatures, while also running a lightweight blockchain node. Our anomaly-based detection utilizes a lightweight supervised SVM model~\cite{8946257}, while the signature-based detection works by matching observed events with the intrusion signatures. The trust management layer consists of smart contracts, which are executed on the blockchain to provide transparent incentive and penalty mechanisms for positive and negative contributions. The storage layer stores the supervised detection models and intrusion signatures in a peer-to-peer hypermedia protocol, IPFS\footnote{\url{https://ipfs.io}}, while the underlying blockchain layer acts as a trusted meta-data storage and consensus platform. We separate the data storage to prevent blockchain size from becoming too large. To link meta-data and the detection model, we store the hash of each model securely in the blockchain.
    
    An intrusion detection module, which is deployed in each IDS node, interacts with blockchain's smart contract to collaboratively contribute a new detection model or to obtain newly published models from other peers. In each interaction, the corresponding smart contract assesses the contribution of the IDS node.
    To avoid an adversary from contributing deceptive detection models, i.e., adversarial attack, the consensus algorithm inspects each contribution for malicious models and exclude the corresponding models from being appended into the blockchain.
    When an IDS node needs to acquire a new detection model, the IDS node collects the model from the storage layer. To notify all peers in case of a detected intrusion, a blockchain event will be triggered. This way, the blockchain event disseminates intrusion alarms to all participating IDS nodes. In our architecture, IDS nodes are placed in gateways, as the IDS nodes are required to perform asymmetric cryptography and machine learning operations, and may need relatively large storage for hosting the blockchain and detection models.
    
    \subsection{Evaluation}
\begin{figure}[!t]
    \centering
    \includegraphics[width=0.4\textwidth]{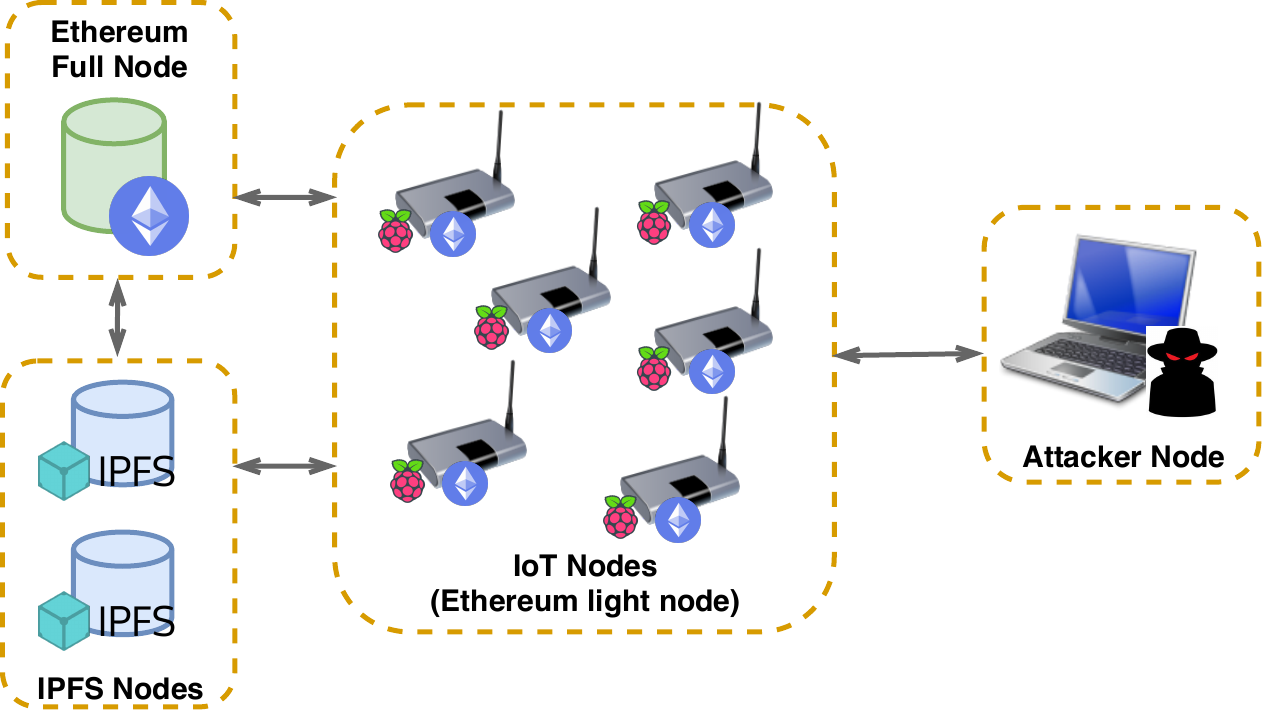}
    \caption{Evaluation scenario.}
    \label{fig:implementation-evaluation}
    \end{figure}

    We are currently implementing a proof-of-concept of our proposed architecture in a local test-bed that simulates real-world IoT deployment, based on the Ethereum blockchain. We show our proof-of-concept design in Fig.~\ref{fig:implementation-evaluation}. Our proof-of-concept consists of four interconnected components, namely IoT networks, an Ethereum full node, an IPFS local cluster, and an attacker simulating incoming attacks. An IoT network of Raspberry Pis running an Ethereum light client and Python based supervised learning modules is implemented as the application layer, which is responsible for detecting and alerting intrusions. We provide a cluster of local IPFS network to store the detection model and datasets, while another instance of Ethereum full node stores the entire blockchain ledger.
    
    For performance evaluation, we are evaluating the effectiveness of our design by penetrating the network with common attacks in IoT network~\cite{Hodo2016}, such as denial of service (DoS), man-in-the-middle or spoofing, reconnaissance, and replay attacks. We also investigate the associated overhead of blockchain utilization in the architecture compared with conventional CIDS mechanisms.
    
\bibliographystyle{./bibliography/IEEEtran}
\bibliography{./bibliography/mybib}

\end{document}